\documentclass[aps,prl,superscriptaddress,twocolumn,amsmath,amssymb,showpacs,floatfix]{revtex4-1}

\usepackage{times}

\usepackage[utf8x]{inputenc}
\usepackage[english]{babel}
\usepackage[T1]{fontenc}
\usepackage{lmodern}
\usepackage{bbold}
\usepackage[pdftex]{graphicx}
\usepackage{epstopdf}

\usepackage{bm}
\usepackage{color}

\newcommand{\bra}[1] {\left \langle #1 \right | }
\newcommand{\ket}[1] { \left | #1 \right \rangle }
\newcommand{\quadra}[1] { \left [ #1 \right ] }
\newcommand{\tonda}[1] { \left ( #1 \right ) }
\newcommand{\graffa}[1] { \left \{ #1 \right \} }

\DeclareMathOperator{\Tr}{Tr}

\begin{document}

\title{Microscopic modelling of general time-dependent quantum Markov processes}

\author{Giulio Amato}
\email{giulio.amato@physik.uni-freiburg.de}
\affiliation{Physikalisches Institut, Universit\"at Freiburg,
  Hermann-Herder-Stra{\ss}e 3, D-79104 Freiburg, Germany}

\author{Heinz-Peter Breuer}
\email{breuer@physik.uni-freiburg.de}
\affiliation{Physikalisches Institut, Universit\"at Freiburg,
  Hermann-Herder-Stra{\ss}e 3, D-79104 Freiburg, Germany}
\affiliation{Freiburg Institute for Advanced Studies (FRIAS), Universität Freiburg,
Albertstraße 19, D-79104 Freiburg, Germany}

\author{Bassano Vacchini}
\affiliation{Dipartimento di Fisica ``Aldo Pontremoli'', Universit{\`a} degli Studi di Milano, via Celoria 16, 20133 Milan, Italy}
\affiliation{Istituto Nazionale di Fisica Nucleare, Sezione di Milano, via Celoria 16, 20133 Milan, Italy}
\email{bassano.vacchini@mi.infn.it}

\begin{abstract}
  Master equations are typically adopted to describe the dynamics of
  open quantum systems. Such equations are either in integro-differential or in
time-local form, with the latter class more frequently adopted due to
  the simpler numerical methods developed to obtain the corresponding
  solution.  Here we show that any time-local master equation with
  positive rates in the generator, i.e. any CP-divisible quantum
  process, admits a microscopic model whose reduced dynamics
  is well described by the given equation.
\end{abstract}

\maketitle

\paragraph{Introduction.} 
In the theory of open quantum systems, master equations are widely
adopted to describe the reduced dynamics. Such equations are typically
derived from the full system-environment evolution via projection
operator techniques, once an average over the environmental degrees
of freedom is performed, and can be divided in two categories:
integro-differential and time-local equations, respectively called
Nakajima-Zwanzig and time-convolutionless master equations
\cite{Breuer2002,Nakajima1958,Zwanzig1960,Uchiyama1999a,Breuer2006a,Breuer2007a}.
Recently, various results have been obtained for the characterization
of memory kernels leading to well-defined integro-differential quantum
evolution equations
\cite{Budini2006a,Breuer2008a,Vacchini2013a,Chruscinski2016a,Vacchini2016b,Chruscinski2017a},
also pointing to connection with microscopic models
\cite{Budini2013a,Lorenzo2017a}. However, the numerical implementation
of integro-differential equations remains quite demanding
\cite{Montoya-Castillo2016a,Montoya-Castillo2017a}, so that
time-convolutionless master equations often provide a more convenient
approach.  Moreover, in recent years such master equations also attracted a lot
of interest because of their wide use in studying quantum
non-Markovianity \cite{Breuer2012a,Rivas2014a,Breuer2016a}.  In fact, if one considers master equations with the
same operator structure as the so-called
Gorini-Kossakowski-Sudarshan-Lindblad generator
\cite{Gorini1976a,Lindblad1976a}, which warrants hermiticity and trace
preservation of the statistical operator, the relationship between the
rates and the Lindblad operators allows to assess the divisibility
character of the time evolution. Namely, whether the evolution map can
be written as composition of maps referring to arbitrary intermediate
time intervals. In particular, if the time evolution can be expressed
as a composition of completely positive maps, the dynamics is said to
be CP-divisible \cite{Breuer2012a,Breuer2016a}.  In accordance with different recently proposed
definitions and measures of quantum non-Markovianity, this
divisibility enables a characterization of quantum memory effects,
which are absent in the case of a CP-divisible
dynamics \cite{Breuer2009b,Rivas2010a,Chruscinski2011a,Wissmann2015a,Amato2018a}.
The solutions of time-convolutionless master equations characterized
by positive rates in the generator, besides describing a well-defined
evolution equation, do provide CP-divisible time evolutions
\cite{Breuer2012a,Breuer2016a}.  Nonetheless, master equations are not always
derived from an underlying Hamiltonian model, they are often rather
adopted to describe phenomenologically the non unitary dynamics of open
quantum systems. It is hence an important task to clarify whether or
not such equations actually correspond to a well-defined physical model.

The goal of this work is indeed to show that a
  system-environment interaction Hamiltonian can always be associated
  to any CP-divisible master equation, so that the reduced dynamics of
  the open system is described by the given equation.  This is
obtained considering the interaction of the system with multiple
independent bosonic baths, truncated to the Born approximation, in the
limit of infinite spectral densities bandwidth and assuring that there
is no contribution to the two-time environmental correlation function
due to negative frequencies components of the spectral
density. This result shows that Markovian master equations are
  always connected to a modelling via the introduction of a bosonic
  environment and a suitable system-environment interaction Hamiltonian, for
  which they give an accurate description of the reduced dynamics,
  once some approximation, like weak-coupling and separation of
  relevant time-scales between system and environment, are
  adopted. Our work further opens the way to considering actual
  physical models in which such microscopic interaction can be realized.


Given an open quantum system $ S$, with associated Hilbert space $ \mathcal{H}_S $ such that $ \textrm{dim} \mathcal{H}_S = n $, it is possible to describe its non unitary dynamics via the following time-convolutionless master equation
\begin{align}
\label{eq:CPdivMicroModel}&\frac{\textrm{d}}{\textrm{d}t} \rho_S (t) = - \frac{i}{\hbar} [ H_S
  (t) , \rho_S (t) ]
  \\
 &\, + \! \sum_{k=1}^{n^2 - 1} \gamma_k (t) \left[ A_k (t) \rho_S
    (t) A_k^{\dagger} (t) - \frac{1}{2} \{ A_k^{\dagger} (t) A_k (t) ,
    \rho_S (t) \} \right] ,
  \nonumber
\end{align}
where $ \{ A_k (t) \}_k $ are the so-called Lindblad operators and the
real coefficients $ \gamma_k (t) $ are called rates. 
If all rates are positive for $ t \geq 0 $, the associated quantum
process is CP-divisible, according to the definition given above,
that is, it gives rise to an evolution map that can be expressed as the
composition of completely positive maps over any subdivision of the
overall evolution time \cite{Breuer2012a, Breuer2016a}.  In particular, positivity of the rates
guarantees complete positivity of the reduced dynamics.



The main result of this contribution is to present a microscopic
model, whose reduced dynamics is described, in a suitable limit, by a
master equation of the form \eqref{eq:CPdivMicroModel}.

\begin{figure}[h]
\centering
\includegraphics[width=.48\textwidth]{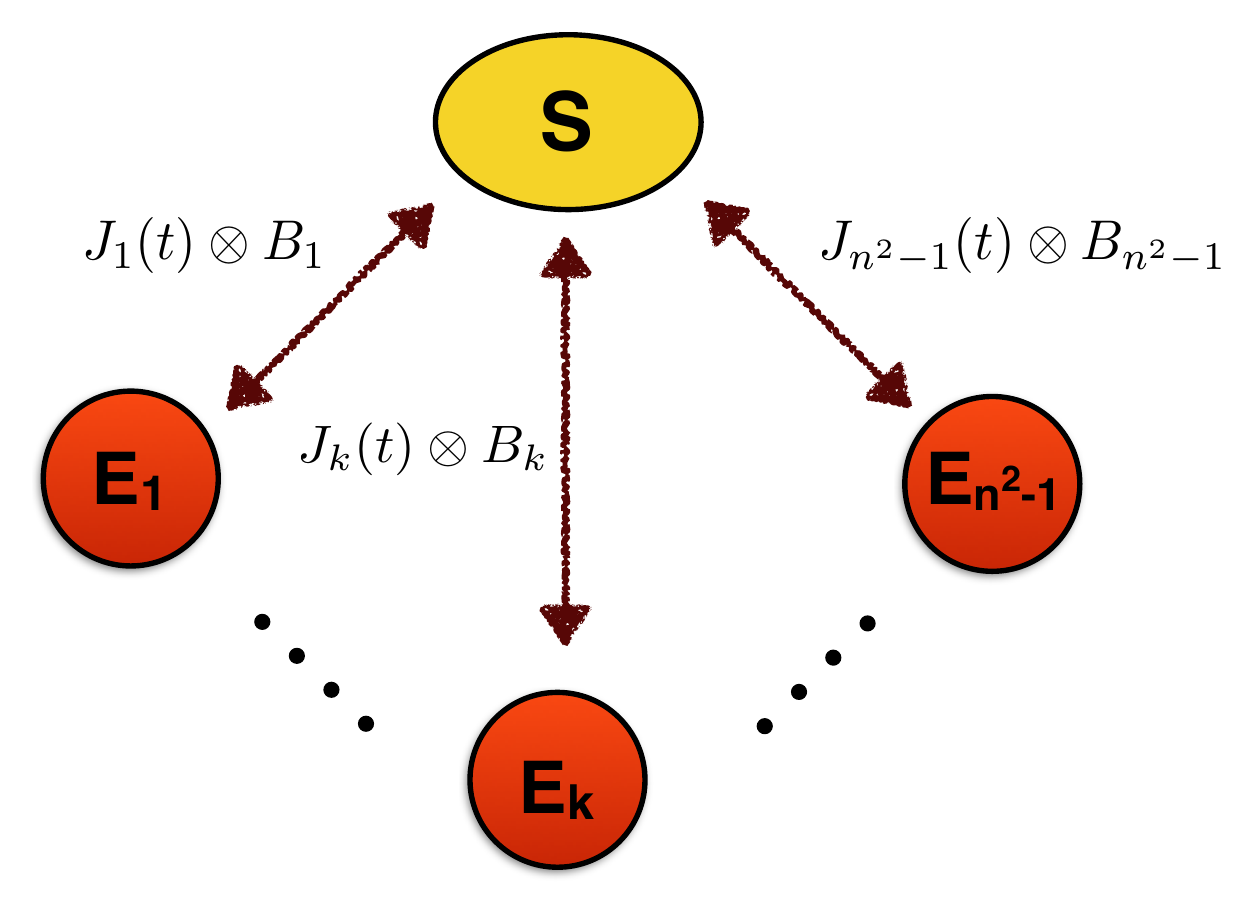}
\caption{Schematic representation of the considered microscopic model,
  in which the system $ S $ interacts with $n^2-1$ independent bosonic
  environments $ E_k $, each corresponding to a dissipation channel in the
  master equation. The coupling is given through the dressed
  Lindblad operators $ J_k (t)=\sqrt{{\gamma_k(t)}/{\gamma_0}}A_k(t)$
  of Eq.~\eqref{eq:1} and the combination of field operators $ B_k$ as given in Eq.~\eqref{eq:conjOpEnv}.}
\label{fig:modelCPDiv2}
\end{figure}

\paragraph{Microscopic model.} We let the system $ S$ interact with
independent bosonic baths $ E_k $, as depicted in
Fig.~\ref{fig:modelCPDiv2}, considering as an Ansatz the following
total system-environment  Hamiltonian
\begin{equation}
H (t) = \textrm{ } H_S (t) + H_E + H_{\textrm{int}} (t), 
\label{eq:HamGenCPDivisib}
\end{equation}
with $ H_S (t) $ the Hamiltonian appearing in the commutator term
in Eq.~\eqref{eq:CPdivMicroModel}, while $ H_E $ is the free Hamiltonian
of the independent bosonic baths, so that
\begin{equation}
  \label{eq:3}
  H_E = \textrm{ } \sum_{k=1}^{n^2 - 1} H_{E_k} 
\end{equation}
with
\begin{equation}
H_{E_k} = \sum_{\mu} \hbar ( \omega_{k, \mu} - \omega_0 ) b^{\dagger}_{k, \mu} b_{k, \mu} ,
\label{eq:structBosE}
\end{equation} 
where $ b_{k, \mu}$ and $b^{\dagger}_{k, \mu}$ denote
creation and annihilation bosonic operators satisfying the canonical commutation relations
\begin{equation}
\label{eq:commStructBos}
  [ b_{k, \mu}, b_{k', \nu} ] = [ b^{\dagger}_{k, \mu},
  b^{\dagger}_{k', \nu} ] = 0 , \ 
[ b_{k, \mu}, b^{\dagger}_{k', \nu} ] = \delta_{k k'} \delta_{\mu \nu} ,
\end{equation}
and the associated frequencies are taken with respect to a reference
frequency $\omega_0$, resonant with the system \cite{Gardiner2000a}.
The interaction Hamiltonian is taken to be of the form
\begin{equation}
H_{\textrm{int}} (t) = \sum_{k=1}^{n^2 - 1}  [ J_{k} (t) \otimes B^{\dagger}_{k} + J_{k}^{\dagger} (t) \otimes B_k ] ,
\label{eq:conjInteracHam}
\end{equation}
where the system operators  $ \{ J_k (t)
\}_k $  appearing in the coupling term
are related to the original Lindblad operators according to
\begin{equation}
  \label{eq:1}
  J_k (t)=\sqrt{{\gamma_k(t)}/{\gamma_0}}A_k(t),
\end{equation}
where thanks to their positivity the rates, renormalized to a
reference rate $\gamma_0$ which fixes the interaction strength, have been absorbed
in the coupling operators.
Note in particular that we do not put any constraint on the coupling
  strength, we only require that the rates $\gamma_k(t)$ are bounded functions of $t$ within the considered 
time interval and that the Lindblad operators $A_k(t) $ are bounded.
The environmental coupling operators are instead
given by
\begin{equation}
 B_k = \sum_{\mu} g_{k, \mu} b_{k, \mu},
\label{eq:conjOpEnv}
\end{equation}
with $ g_{k, \mu} $ coupling constants. The operators $J_k (t)$ are
thus dimensionless, while $ B_k (t) $
have the dimension of an inverse of time in units such that $ \hbar = 1 $. 
We now move to the interaction picture with respect to the free
Hamiltonian by means of the unitary transformation
\begin{align}
V (t) &=  T_{\leftarrow} \exp \graffa{ -i \int_0^t \textrm{d} \tau \textrm{ } [ H_S (\tau) + H_E  ] } \\
&= {T_{\leftarrow}} \exp \graffa{ - i \int_0^t \textrm{d} \tau
  \textrm{ } H_S (\tau) } \otimes \exp \graffa{ - i H_E t } ,
\nonumber
\label{eq:freeOpUnit}
\end{align}
which can be written $V (t) = {V}_S (t) \otimes {V}_E (t)$, since $ [
H_S (t) , H_E ] = 0 $, and where $ T_{\leftarrow} $ denotes operator
time-ordering. Finally, the interaction Hamiltonian becomes
\begin{equation}
\tilde{H}_{\textrm{int}} (t) =  \sum_{k=1}^{n^2 - 1} \left[ \tilde{J}_{k} (t) \otimes \tilde{B}^{\dagger}_{k} (t) + \tilde{J}_{k}^{\dagger} (t) \otimes \tilde{B}_k (t) \right] ,
\label{eq:conjInteracHamIntPic}
\end{equation}
where the operators in the interaction
picture are indicated by a tilde, so that
$ \tilde{J}_k (t) = {V}^{\dagger}_S (t) J_k (t) {V}_S (t) $
together with $\tilde{B}_k (t) = {V}^{\dagger}_E (t) B_k {V}_E (t) $ which, exploiting
Eq.~(\ref{eq:commStructBos}), takes the form
\begin{equation}
\tilde{B}_k (t) = \sum_{\mu} g_{k, \mu} e^{- i ( \omega_{k, \mu} - \omega_{0 } ) t }  b_{k, \mu} .
\label{eq:BInt}
\end{equation}
The exact evolution of the composite system in the interaction
  picture is thus described by the von Neumann equation
\begin{equation}
\frac{\textrm{d}}{\textrm{d}t} \tilde{\rho} (t) = - i \quadra{ \tilde{H}_{\textrm{int}} (t), \tilde{\rho} (t) } \equiv \mathcal{L} (t) \tilde{\rho} (t),
\label{eq:vonneumannintpicture}
\end{equation}
and in the case of a factorized initial condition
\begin{equation}
\rho (0) = \rho_{S} (0) \otimes \sigma_E 
\label{eq:inCond}
\end{equation}
warrants complete positivity of the reduced dynamics. The state of the environment $ \sigma_E$ is taken 
to be the  tensor product of zero temperature states for each bosonic bath, namely
 $ \sigma_E = \bigotimes_{k=1}^{n^2 - 1} \sigma_{E_k} $ where
 \begin{equation}
   \label{eq:4}
   \sigma_{E_k} = {\ket{ 0} \bra{ 0}}_{E_k},
 \end{equation}
with $ b_{k, \mu}{\ket{ 0} }_{E_k}=0 \ \forall \mu$.

 \paragraph{Time-local master equation.} We now use the projection operator
 technique to derive a time-convolutionless master equation for the
 reduced dynamics \cite{Breuer2002}.  To this aim, we consider a standard projection
 superoperator acting on a generic state $\omega$ of the total system
 as
\begin{equation}
 \mathcal{P } \omega = \Tr_E \omega \otimes \sigma_E,
\label{eq:standardprojection}
\end{equation}
with $ \sigma_E $ the state of the environment appearing in the initial
condition \eqref{eq:inCond}, so that in particular $\mathcal{P } \quadra{ \rho (0) } =  \rho (0)$.
Using the Born approximation, namely considering terms up to the
second order in the interaction Hamiltonian, starting from Eq.~\eqref{eq:vonneumannintpicture}
one obtains
\begin{equation}
 \frac{\textrm{d}}{\textrm{d}t} \mathcal{P} \tilde{\rho} (t) = \int_{0}^t \textrm{d}s \textrm{ } \mathcal{P L} (t) \mathcal{L} (s ) \mathcal{P} \tilde{\rho} (t).  
\label{eq:globalRed}
\end{equation}
Tracing over the degrees of freedom of the environment, by means of a
change of integration variable, one
easily obtains the master equation 
\begin{equation}
 \frac{\textrm{d}}{\textrm{d}t} \tilde{\rho}_{S} (t) =  - \int_{0}^t \textrm{d} \tau \Tr_E [ \tilde{H}_{\textrm{int}} (t), [ \tilde{H}_{\textrm{int}} (t- \tau ) , \tilde{\rho}_{S} (t) \otimes \sigma_E ]] ,
\label{eq:redfield}
\end{equation}
which is sometimes called Redfield equation.
We further expand the commutators and use the identities,
valid $ \forall\ i, k = 1,..., n^2 - 1 $ and $ \forall\ t,s \geq 0 $
\begin{equation}
\begin{aligned}
\Tr_E \{ \tilde{B}_i (t) \tilde{B}_k (s) \sigma_E \} &=  \Tr_E
\{ \tilde{B}^{\dagger}_i (t) \tilde{B}^{\dagger}_k (s) \sigma_E
\} =  0,
 \\
\Tr_E \{ \tilde{B}^{\dagger}_i (t) \tilde{B}_k (s) \sigma_E \}  &=  0,
\end{aligned}
\end{equation}
together with
\begin{equation}
  \Tr_E \{ \tilde{B}_i (t) \tilde{B}^{\dagger}_k (s) \sigma_E \} =  0 \label{eq:crossless}
\end{equation}
for $i\not = k$. 
The only nontrivial two-time correlation function is thus given by
\begin{equation}
\begin{aligned}
  \label{eq:2}
  \Tr_E \{
\tilde{B}_k (t) \tilde{B}^{\dagger}_k (s) \sigma_E \}
&=
\Tr_{E_k} \{ \tilde{B}_{k} (t ) \tilde{B}^{\dagger}_{k} (s )
                                                      \sigma_{E_k}
                                                      \} 
\\
& \equiv \langle \tilde{B}_{k} (t ) \tilde{B}^{\dagger}_{k} (s
  ) \rangle
\end{aligned}
\end{equation}
so that, also
exploiting the cyclic property of the partial trace, we finally obtain
\begin{equation}
\frac{\textrm{d}}{\textrm{d}t} \tilde{\rho}_{S} (t) =\sum_{k=1}^{n^2-1} \int_{0}^t \textrm{d} \tau \mathcal{K}_k(t,\tau)[ \tilde{\rho}_{S} (t)  ] ,
\label{eq:preFinMicro}
\end{equation}
with
\begin{displaymath}
\begin{aligned}
\mathcal{K}_k(t,\tau)[ \tilde{\rho}_{S} (t)  ] \!=\! &
 \textrm{ } \tilde{J}_{k} (t- \tau ) \tilde{\rho}_{S} (t)  \tilde{J}^{\dagger}_{k} (t )  \langle \tilde{B}_{k} (t ) \tilde{B}^{\dagger}_{k} (t - \tau ) \rangle  \\ 
&+ \textrm{ } \tilde{J}_{k} (t ) \tilde{\rho}_{S} (t) \tilde{J}^{\dagger}_{k} (t- \tau ) \langle \tilde{B}_{k} ( t - \tau  )  \tilde{B}^{\dagger}_{k} (t ) \rangle   \\
 &- \textrm{ } \tilde{J}^{\dagger}_{k} ( t ) \tilde{J}_{k} ( t- \tau ) \tilde{\rho}_{S} (t) \langle \tilde{B}_{k} (t ) \tilde{B}^{\dagger}_{k} (t - \tau ) \rangle  \\
&-  \textrm{ } \tilde{\rho}_{S} (t) \tilde{J}^{\dagger}_{k} (t- \tau ) \tilde{J}_{k} (t ) \langle \tilde{B}_{k} (t - \tau )  \tilde{B}^{\dagger}_{k} (t  ) \rangle  .
\end{aligned}
\end{displaymath}
We stress that no secular approximation is involved in obtaining
  this expression, while the
  absence of cross terms in the index $k$ is due to Eq.~\eqref{eq:crossless}.
In particular, recalling the expression of the environmental states
[Eq.~\eqref{eq:4}] we have
\begin{equation}
\label{eq:7}
\begin{aligned}
 \langle \tilde{B}_k (t) \tilde{B}^{\dagger}_k (s) \rangle = & \bra{
                                                               0}
                                                               \tilde{B}_k
                                                               (t)
                                                               \tilde{B}^{\dagger}_k
                                                               (s)
                                                               \ket{
                                                               0}  \\
 = & \sum_{\mu} \vert g_{k, \mu} \vert^2 e^{- i ( \omega_{k, \mu} - \omega_{0} ) (t - s ) }. 
\end{aligned}
\end{equation}
If we now consider a continuum of environmental modes
characterized by a given density of states, thus replacing the sum
over $\mu$ weighted by the coupling constants $g_{k, \mu}$ with an
integral over $\omega$ with a suitable spectral density  $I_k (\omega)$, these correlation
functions can be expressed according to \cite{Breuer2002} in the form
\begin{equation}
 \langle \tilde{B}_k (t) \tilde{B}^{\dagger}_k (s) \rangle = \int_0^{+\infty} \textrm{d} \omega \textrm{ } I_k ( \omega) e^{ -i ( \omega - \omega_{0} ) (t-s) }. 
\label{eq:2pCorrF}
\end{equation}
We now consider all spectral densities to be proportional to the same Cauchy-Lorentz distribution
\begin{equation}
I_k ( \omega) =I ( \omega) = \frac{1}{2 \pi } \frac{ \gamma_0 \lambda^2 }{ (\omega - \omega_{0} )^2 + \lambda^2  } , 
\label{eq:spectDensCauchy}
\end{equation}
with resonant frequency $ \omega_0 $ corresponding to the
reference frequency considered in Eq.~(\ref{eq:structBosE}) and a
common factor $\gamma_0$ given by the reference rate introduced in
Eq.~ (\ref{eq:1}). The spectral width $ \lambda $ in
Eq.~(\ref{eq:spectDensCauchy}) 
is connected to the typical environmental correlation time $ \tau_E $ by the relation
\begin{equation}
\tau_E = \lambda^{-1} ,
\label{eq:relImp}
\end{equation}
while the typical relaxation time scale for the system is set by
  $\tau_R=\gamma_0^{-1}$. As shown in \cite{Breuer2002} for this
expression of the spectral density the Born approximation considered
in Eq.~\eqref{eq:standardprojection} is indeed justified if
$\gamma_0 / \lambda \ll 1$, corresponding to a separation of
time scales accounting for a Markovian dynamics.
We stress that in our derivation $\gamma_k(t)$ and $A_k(t)$ are fixed by
  the master equation Eq.~\eqref{eq:CPdivMicroModel}, while we can freely
choose the parameters $\gamma_0$ and $\lambda$ in order to satisfy the
constraints ${{\gamma_k(t)}/{\gamma_0}}\lessapprox 1$ and $\gamma_0 / \lambda \ll 1$.
Due to its shape the spectral density in
Eq.~\eqref{eq:spectDensCauchy} is related, via Fourier transform, to
an exponential decaying function of time, namely
\begin{equation}
\int_{-\infty}^{+\infty} \textrm{d} \omega \textrm{ } I ( \omega )
e^{- i ( \omega - \omega_0 ) \tau } = \frac{\gamma_0 \lambda}{2} \exp ( - \lambda \vert \tau \vert  ). 
\label{eq:spectDensCauchyprop}
\end{equation}
Nevertheless,  an exact calculation calls for an integration over
positive physical frequencies only, so that Eq.~(\ref{eq:2pCorrF}) can be
written as
\begin{equation}
\int_0^{+\infty} \textrm{d} \omega I ( \omega) e^{ -i ( \omega - \omega_0 ) \tau } =\frac{\gamma_0 \lambda}{2} \exp ( - \lambda \vert \tau \vert  )- R_{\omega_0} (\tau ),
\label{eq:posAxis}
\end{equation}
with 
\begin{equation}
R_{\omega_0} (\tau )  \equiv \int_{- \infty }^0 \textrm{d} \omega  \textrm{ } I (\omega) e^{ -i ( \omega - \omega_0 ) \tau } .
\label{eq:restoCauchy}
\end{equation}
 The latter contribution, however, can be shown to vanish in the limit
$\omega_0 \rightarrow \infty$, to be understood as $\omega_0
  \gg \lambda$. As a general argument, we have that $I (\omega)$ 
is integrable over the negative real axis for $\omega_0 >0$, and
$ | I (\omega) | \xrightarrow{ \omega_0 \rightarrow \infty }  \textrm{
} 0 $. In particular
\begin{equation}
| I (\omega) | \leq \frac{1}{2 \pi } \frac{ \gamma_0 \lambda^2}{(\omega - \Omega )^2 } = g (\omega)
\end{equation}
with $\Omega$ a positive frequency strictly smaller than $\omega_0$,
i.e. $0<\Omega<\omega_0$. We can thus apply the dominated convergence
theorem with respect to the dominating function $g (\omega)$ to conclude
\begin{equation}
\vert R_{\omega_0} (\tau ) \vert \xrightarrow{ \omega_0 \rightarrow \infty } 0 
\label{eq:restoCauchyZero}
\end{equation}
and, therefore, in the same limit,
\begin{equation}
  \label{eq:5}
  \int_0^{+\infty} \textrm{d} \omega \textrm{ } I ( \omega) e^{ -i ( \omega - \omega_0 ) \tau } = \frac{1}{2} \gamma_0 \lambda \exp ( - \lambda \vert \tau \vert  ).
\end{equation}
Note that in this limit the correlation function
  Eq.~(\ref{eq:2pCorrF}) becomes purely real, so that no Lamb shift
  correction term appears.
  A closer inspection of $R_{\omega_0} (\tau )$ for the case of
a Lorentzian spectral density, as considered in \cite{Khalfin1958a},
indeed shows that assuming for simplicity $\tau
\sqrt{\omega_0^2+\lambda^2}\gg 1$, the neglected contribution reads
\begin{equation}
  \label{eq:Khalfin}
  R_{\omega_0}(\tau) = \frac{i}{2\pi} \frac{\gamma_0}{\tau} \frac{\lambda^2}{\omega_0^2+\lambda^2} 
 e^{i\omega_0\tau}.
\end{equation}
In particular in order to take $R_{\omega_0}$ to be small one has to take
the limit $\omega_0\to\infty$ before considering large $\lambda$. If also the bandwidth of the spectral density goes to
infinity, i.e. $ \lambda \rightarrow \infty $, the two-time
environmental correlation function is proportional to a Dirac delta function, namely
\begin{equation}
 \langle \tilde{B}_k (t) \tilde{B}^{\dagger}_k (t-\tau ) \rangle = \frac{\gamma_0 \lambda}{2} \exp ( - \lambda \vert \tau \vert  ) 
\xrightarrow{\lambda \to \infty}  \gamma_0 \delta ( \tau ) .
\label{eq:cauchydelta}
\end{equation}
The use of this limit in Eq.~(\ref{eq:preFinMicro}) is justified if
the  typical time scale $\tau_J$ for the change of the
operators $\tilde{J}_k(t)$ appearing in the expression is much larger than the environmental time
scale as set by  Eq.~(\ref{eq:relImp}), i.e.  $\lambda\gg 1/\tau_J$, leading therefore to the following
relationship for the validity of the different approximations
\begin{equation}
  \label{eq:summary}
  \omega_0 \gg \lambda \gg \gamma_0, \tau_J^{-1}.
\end{equation}
In this limit Eq.~(\ref{eq:preFinMicro}) finally reads
\begin{align}
\frac{\textrm{d}}{\textrm{d}t} &\tilde{\rho}_S (t) =\\ &\!\!\!\! \sum_{k=1}^{n^2 -
  1} \gamma_k (t) \left[ \tilde{A}_k (t) \tilde{\rho}_{S} (t) \tilde{A}_k^{\dagger} (t) - \frac{1}{2} \graffa{ \tilde{A}_k^{\dagger} (t) \tilde{A}_k (t), \tilde{\rho}_{S} (t) } \right], \nonumber
\label{eq:CPdivMicroModelInt}
\end{align}
where the time dependent rates $\gamma_k (t)$ are exactly those
appearing in the original time-convolutionless  generator
Eq.~(\ref{eq:CPdivMicroModel}), which is recovered switching back again to the Schr\"{o}dinger picture.

%
%

\paragraph{Discussion.} 

We note that the considered proof does not actually rely on the
specific choice of Cauchy-Lorentz spectral density in
Eq.~\eqref{eq:spectDensCauchy}. Only two general requirements have to
be satisfied by the considered spectral density, namely: \emph{i}) the integration of the spectral density over
negative frequencies in the limit $ \omega_0 \rightarrow \infty $
should provide a
vanishing contribution; \emph{ii})
the two-time correlation function of the
environment obtained from the Fourier transform of the spectral
density, in the limit of infinite bandwidth  $ \lambda \rightarrow \infty $, should behave as a Dirac delta function.  Examples of
spectral densities which satisfy such conditions, for which an
analogous calculation can be performed, are e.g.  Gaussian spectral
densities
\begin{equation}
I ( \omega  ) = \frac{\gamma_0 \lambda}{ \sqrt{ 2 } } e^{ - \frac{  (\omega - \omega_{0} )^2} { 2 \lambda^2  } } ,
\label{eq:gaussianspd}
\end{equation} and
spectral densities given by the squared sinc
\begin{equation}
 I ( \omega ) = \frac{\gamma_0}{ 2\pi }\textrm{ sinc}^2 \tonda{ \frac{\omega - \omega_0 }{ \lambda } },
\label{eq:squaresincspd}
\end{equation}
as well as as general linear combination of spectral densities with these properties.

It is well known that derivations of a master equation in Lindblad
form, for the case of constant rates, can be obtained along different
paths, see e.g. \cite{Alicki2002b} for a review. One can
consider suitable mathematical limits once the environmental degrees
of freedom have been traced over
\cite{Davies1974a,Gorini1976b,Dumcke1985a}. In a different formalism,
one can obtain the reduced
dynamics in Lindblad form as an exact result once suitable
approximations have been introduced at the level of the coupling term
as in quantum stochastic calculus
\cite{Hudson1984a,Barchielli2006a,Barchielli2013a,Barchielli2015a}. In
this paper we have shown that also for the case of non-constant but
positive rates a microscopic model can be pointed out that leads in
suitable limits to the desired master equation. In particular, we have
considered a weak-coupling situation and spectral densities such that
conditions corresponding to the flat spectrum and broadband
approximation used in standard derivations apply
\cite{Gardiner2000a}. Alternatively
one might consider the continuous limit of memoryless collision
models, which presently has only been used for the derivation of the standard
Lindblad master equation with constant coefficients \cite{Ziman2005a,Ciccarello2017a}.

\paragraph{Conclusions.} We have thus shown that any master
  equation in time-local form with positive rates can be connected to
  a system-environment model, considering a suitable Ansatz for the
  interaction Hamiltonian. The proof relies on a microscopic model,
where the system of interest interacts with multiple independent
bosonic baths, each in the ground state, for which the master equation
of interest provides an accurate description of the reduced
dynamics. Our construction highlights the relationship between
Markovianity of the generated process and typical limits considered in
the literature, such as weak coupling and separation of system and
environment time scales.  The development of other approaches, such as
  quantum stochastic calculus or, in a less rigorous framework,
  collision models might lead to different insights.
  We stress however that further work is anyhow needed
  to connect the considered system-environment interaction Hamiltonian
  with a specific physical implementation.

Note moreover that a crucial ingredient of our construction is the assumption
of the positivity of the rates in the master equation of interest,
which are then absorbed in the Lindblad operators. This prevents the
possibility to extend this construction, as well as similar ones, to
master equations with negative rates and, hence, a novel approach is
necessary to connect master equations in time-local form with rates
that can take on negative values with an underlying microscopic
interaction model.  Indeed, when one allows for negativity of the
coefficients appearing in the master equation, even complete
positivity of the obtained evolution is not granted.  The
characterization of the most general structure of time-convolutionless
master equation admitting as solutions well-defined, i.e. completely
positive trace preserving maps, remains an important open
problem.

\paragraph{Acknowledgments.}
GA would like to thank the the German Research Foundation (DFG) and Fondazione Grazioli for support. HPB
and BV acknowledge support from the Joint Project 
"Quantum Information Processing in Non-Markovian Quantum Complex
Systems" funded by FRIAS/University of Freiburg and IAR/Nagoya
University. BV further
acknowledges support from the FFABR project of MIUR.


\begin{thebibliography}{38}%
\makeatletter
\providecommand \@ifxundefined [1]{%
 \@ifx{#1\undefined}
}%
\providecommand \@ifnum [1]{%
 \ifnum #1\expandafter \@firstoftwo
 \else \expandafter \@secondoftwo
 \fi
}%
\providecommand \@ifx [1]{%
 \ifx #1\expandafter \@firstoftwo
 \else \expandafter \@secondoftwo
 \fi
}%
\providecommand \natexlab [1]{#1}%
\providecommand \enquote  [1]{``#1''}%
\providecommand \bibnamefont  [1]{#1}%
\providecommand \bibfnamefont [1]{#1}%
\providecommand \citenamefont [1]{#1}%
\providecommand \href@noop [0]{\@secondoftwo}%
\providecommand \href [0]{\begingroup \@sanitize@url \@href}%
\providecommand \@href[1]{\@@startlink{#1}\@@href}%
\providecommand \@@href[1]{\endgroup#1\@@endlink}%
\providecommand \@sanitize@url [0]{\catcode `\\12\catcode `\$12\catcode
  `\&12\catcode `\#12\catcode `\^12\catcode `\_12\catcode `\%12\relax}%
\providecommand \@@startlink[1]{}%
\providecommand \@@endlink[0]{}%
\providecommand \url  [0]{\begingroup\@sanitize@url \@url }%
\providecommand \@url [1]{\endgroup\@href {#1}{\urlprefix }}%
\providecommand \urlprefix  [0]{URL }%
\providecommand \Eprint [0]{\href }%
\providecommand \doibase [0]{http://dx.doi.org/}%
\providecommand \selectlanguage [0]{\@gobble}%
\providecommand \bibinfo  [0]{\@secondoftwo}%
\providecommand \bibfield  [0]{\@secondoftwo}%
\providecommand \translation [1]{[#1]}%
\providecommand \BibitemOpen [0]{}%
\providecommand \bibitemStop [0]{}%
\providecommand \bibitemNoStop [0]{.\EOS\space}%
\providecommand \EOS [0]{\spacefactor3000\relax}%
\providecommand \BibitemShut  [1]{\csname bibitem#1\endcsname}%
\let\auto@bib@innerbib\@empty
\bibitem [{\citenamefont {Breuer}\ and\ \citenamefont
  {Petruccione}(2002)}]{Breuer2002}%
  \BibitemOpen
  \bibfield  {author} {\bibinfo {author} {\bibfnamefont {H.-P.}\ \bibnamefont
  {Breuer}}\ and\ \bibinfo {author} {\bibfnamefont {F.}~\bibnamefont
  {Petruccione}},\ }\href@noop {} {\emph {\bibinfo {title} {The Theory of Open
  Quantum Systems}}}\ (\bibinfo  {publisher} {Oxford University Press},\
  \bibinfo {address} {Oxford},\ \bibinfo {year} {2002})\BibitemShut {NoStop}%
\bibitem [{\citenamefont {Nakajima}(1958)}]{Nakajima1958}%
  \BibitemOpen
  \bibfield  {author} {\bibinfo {author} {\bibfnamefont {S.}~\bibnamefont
  {Nakajima}},\ }\href@noop {} {\bibfield  {journal} {\bibinfo  {journal}
  {Progr. Theor. Phys.}\ }\textbf {\bibinfo {volume} {20}},\ \bibinfo {pages}
  {948} (\bibinfo {year} {1958})}\BibitemShut {NoStop}%
\bibitem [{\citenamefont {Zwanzig}(1960)}]{Zwanzig1960}%
  \BibitemOpen
  \bibfield  {author} {\bibinfo {author} {\bibfnamefont {R.}~\bibnamefont
  {Zwanzig}},\ }\href@noop {} {\bibfield  {journal} {\bibinfo  {journal} {J.
  Chem. Phys.}\ }\textbf {\bibinfo {volume} {33}},\ \bibinfo {pages} {1338}
  (\bibinfo {year} {1960})}\BibitemShut {NoStop}%
\bibitem [{\citenamefont {Uchiyama}\ and\ \citenamefont
  {Shibata}(1999)}]{Uchiyama1999a}%
  \BibitemOpen
  \bibfield  {author} {\bibinfo {author} {\bibfnamefont {C.}~\bibnamefont
  {Uchiyama}}\ and\ \bibinfo {author} {\bibfnamefont {F.}~\bibnamefont
  {Shibata}},\ }\href {\doibase 10.1103/PhysRevE.60.2636} {\bibfield  {journal}
  {\bibinfo  {journal} {Phys. Rev. E}\ }\textbf {\bibinfo {volume} {60}},\
  \bibinfo {pages} {2636} (\bibinfo {year} {1999})}\BibitemShut {NoStop}%
\bibitem [{\citenamefont {Breuer}\ \emph {et~al.}(2006)\citenamefont {Breuer},
  \citenamefont {Gemmer},\ and\ \citenamefont {Michel}}]{Breuer2006a}%
  \BibitemOpen
  \bibfield  {author} {\bibinfo {author} {\bibfnamefont {H.-P.}\ \bibnamefont
  {Breuer}}, \bibinfo {author} {\bibfnamefont {J.}~\bibnamefont {Gemmer}}, \
  and\ \bibinfo {author} {\bibfnamefont {M.}~\bibnamefont {Michel}},\
  }\href@noop {} {\bibfield  {journal} {\bibinfo  {journal} {Phys. Rev.~E}\
  }\textbf {\bibinfo {volume} {73}},\ \bibinfo {eid} {016139} (\bibinfo {year}
  {2006})}\BibitemShut {NoStop}%
\bibitem [{\citenamefont {Breuer}(2007)}]{Breuer2007a}%
  \BibitemOpen
  \bibfield  {author} {\bibinfo {author} {\bibfnamefont {H.-P.}\ \bibnamefont
  {Breuer}},\ }\href@noop {} {\bibfield  {journal} {\bibinfo  {journal} {Phys.
  Rev. A}\ }\textbf {\bibinfo {volume} {75}},\ \bibinfo {eid} {022103}
  (\bibinfo {year} {2007})}\BibitemShut {NoStop}%
\bibitem [{\citenamefont {Budini}(2006)}]{Budini2006a}%
  \BibitemOpen
  \bibfield  {author} {\bibinfo {author} {\bibfnamefont {A.~A.}\ \bibnamefont
  {Budini}},\ }\href@noop {} {\bibfield  {journal} {\bibinfo  {journal} {Phys.
  Rev.~A}\ }\textbf {\bibinfo {volume} {74}},\ \bibinfo {pages} {053815}
  (\bibinfo {year} {2006})}\BibitemShut {NoStop}%
\bibitem [{\citenamefont {Breuer}\ and\ \citenamefont
  {Vacchini}(2008)}]{Breuer2008a}%
  \BibitemOpen
  \bibfield  {author} {\bibinfo {author} {\bibfnamefont {H.-P.}\ \bibnamefont
  {Breuer}}\ and\ \bibinfo {author} {\bibfnamefont {B.}~\bibnamefont
  {Vacchini}},\ }\href@noop {} {\bibfield  {journal} {\bibinfo  {journal}
  {Phys. Rev. Lett.}\ }\textbf {\bibinfo {volume} {101}},\ \bibinfo {eid}
  {140402} (\bibinfo {year} {2008})}\BibitemShut {NoStop}%
\bibitem [{\citenamefont {Vacchini}(2013)}]{Vacchini2013a}%
  \BibitemOpen
  \bibfield  {author} {\bibinfo {author} {\bibfnamefont {B.}~\bibnamefont
  {Vacchini}},\ }\href@noop {} {\bibfield  {journal} {\bibinfo  {journal}
  {Phys. Rev.~A}\ }\textbf {\bibinfo {volume} {87}},\ \bibinfo {pages}
  {030101(R)} (\bibinfo {year} {2013})}\BibitemShut {NoStop}%
\bibitem [{\citenamefont {Chruscinski}\ and\ \citenamefont
  {Kossakowski}(2016)}]{Chruscinski2016a}%
  \BibitemOpen
  \bibfield  {author} {\bibinfo {author} {\bibfnamefont {D.}~\bibnamefont
  {Chruscinski}}\ and\ \bibinfo {author} {\bibfnamefont {A.}~\bibnamefont
  {Kossakowski}},\ }\href {\doibase 10.1103/PhysRevA.94.020103} {\bibfield
  {journal} {\bibinfo  {journal} {Phys. Rev. A}\ }\textbf {\bibinfo {volume}
  {94}},\ \bibinfo {pages} {020103(R)} (\bibinfo {year} {2016})}\BibitemShut
  {NoStop}%
\bibitem [{\citenamefont {Vacchini}(2016)}]{Vacchini2016b}%
  \BibitemOpen
  \bibfield  {author} {\bibinfo {author} {\bibfnamefont {B.}~\bibnamefont
  {Vacchini}},\ }\href {\doibase 10.1103/PhysRevLett.117.230401} {\bibfield
  {journal} {\bibinfo  {journal} {Phys. Rev. Lett.}\ }\textbf {\bibinfo
  {volume} {117}},\ \bibinfo {pages} {230401} (\bibinfo {year}
  {2016})}\BibitemShut {NoStop}%
\bibitem [{\citenamefont {Chruscinski}\ and\ \citenamefont
  {Kossakowski}(2017)}]{Chruscinski2017a}%
  \BibitemOpen
  \bibfield  {author} {\bibinfo {author} {\bibfnamefont {D.}~\bibnamefont
  {Chruscinski}}\ and\ \bibinfo {author} {\bibfnamefont {A.}~\bibnamefont
  {Kossakowski}},\ }\href {\doibase 10.1103/PhysRevA.95.042131} {\bibfield
  {journal} {\bibinfo  {journal} {Phys. Rev. A}\ }\textbf {\bibinfo {volume}
  {95}},\ \bibinfo {pages} {042131} (\bibinfo {year} {2017})}\BibitemShut
  {NoStop}%
\bibitem [{\citenamefont {Budini}(2013)}]{Budini2013a}%
  \BibitemOpen
  \bibfield  {author} {\bibinfo {author} {\bibfnamefont {A.~A.}\ \bibnamefont
  {Budini}},\ }\href {\doibase 10.1103/PhysRevA.88.032115} {\bibfield
  {journal} {\bibinfo  {journal} {Phys. Rev. A}\ }\textbf {\bibinfo {volume}
  {88}},\ \bibinfo {pages} {032115} (\bibinfo {year} {2013})}\BibitemShut
  {NoStop}%
\bibitem [{\citenamefont {Lorenzo}\ \emph {et~al.}(2017)\citenamefont
  {Lorenzo}, \citenamefont {Ciccarello}, \citenamefont {Palma},\ and\
  \citenamefont {Vacchini}}]{Lorenzo2017a}%
  \BibitemOpen
  \bibfield  {author} {\bibinfo {author} {\bibfnamefont {S.}~\bibnamefont
  {Lorenzo}}, \bibinfo {author} {\bibfnamefont {F.}~\bibnamefont {Ciccarello}},
  \bibinfo {author} {\bibfnamefont {G.~M.}\ \bibnamefont {Palma}}, \ and\
  \bibinfo {author} {\bibfnamefont {B.}~\bibnamefont {Vacchini}},\ }\href
  {\doibase 10.1142/S123016121740011X} {\bibfield  {journal} {\bibinfo
  {journal} {Open Syst. Inf. Dyn.}\ }\textbf {\bibinfo {volume} {24}},\
  \bibinfo {pages} {1740011} (\bibinfo {year} {2017})}\BibitemShut {NoStop}%
\bibitem [{\citenamefont {Montoya-Castillo}\ and\ \citenamefont
  {Reichman}(2016)}]{Montoya-Castillo2016a}%
  \BibitemOpen
  \bibfield  {author} {\bibinfo {author} {\bibfnamefont {A.}~\bibnamefont
  {Montoya-Castillo}}\ and\ \bibinfo {author} {\bibfnamefont {D.~R.}\
  \bibnamefont {Reichman}},\ }\href {\doibase 10.1063/1.4948408} {\bibfield
  {journal} {\bibinfo  {journal} {J. Chem. Phys.}\ }\textbf {\bibinfo {volume}
  {144}},\ \bibinfo {pages} {184104} (\bibinfo {year} {2016})}\BibitemShut
  {NoStop}%
\bibitem [{\citenamefont {Montoya-Castillo}\ and\ \citenamefont
  {Reichman}(2017)}]{Montoya-Castillo2017a}%
  \BibitemOpen
  \bibfield  {author} {\bibinfo {author} {\bibfnamefont {A.}~\bibnamefont
  {Montoya-Castillo}}\ and\ \bibinfo {author} {\bibfnamefont {D.~R.}\
  \bibnamefont {Reichman}},\ }\href {\doibase 10.1063/1.4975388} {\bibfield
  {journal} {\bibinfo  {journal} {J. Chem. Phys.}\ }\textbf {\bibinfo {volume}
  {146}},\ \bibinfo {pages} {084110} (\bibinfo {year} {2017})}\BibitemShut
  {NoStop}%
\bibitem [{\citenamefont {Breuer}(2012)}]{Breuer2012a}%
  \BibitemOpen
  \bibfield  {author} {\bibinfo {author} {\bibfnamefont {H.-P.}\ \bibnamefont
  {Breuer}},\ }\href@noop {} {\bibfield  {journal} {\bibinfo  {journal} {J.
  Phys. B}\ }\textbf {\bibinfo {volume} {45}},\ \bibinfo {pages} {154001}
  (\bibinfo {year} {2012})}\BibitemShut {NoStop}%
\bibitem [{\citenamefont {Rivas}\ \emph {et~al.}(2014)\citenamefont {Rivas},
  \citenamefont {Huelga},\ and\ \citenamefont {Plenio}}]{Rivas2014a}%
  \BibitemOpen
  \bibfield  {author} {\bibinfo {author} {\bibfnamefont {A.}~\bibnamefont
  {Rivas}}, \bibinfo {author} {\bibfnamefont {S.~F.}\ \bibnamefont {Huelga}}, \
  and\ \bibinfo {author} {\bibfnamefont {M.~B.}\ \bibnamefont {Plenio}},\
  }\href@noop {} {\bibfield  {journal} {\bibinfo  {journal} {Rep. Prog. Phys.}\
  }\textbf {\bibinfo {volume} {77}},\ \bibinfo {pages} {094001} (\bibinfo
  {year} {2014})}\BibitemShut {NoStop}%
\bibitem [{\citenamefont {Breuer}\ \emph {et~al.}(2016)\citenamefont {Breuer},
  \citenamefont {Laine}, \citenamefont {Piilo},\ and\ \citenamefont
  {Vacchini}}]{Breuer2016a}%
  \BibitemOpen
  \bibfield  {author} {\bibinfo {author} {\bibfnamefont {H.-P.}\ \bibnamefont
  {Breuer}}, \bibinfo {author} {\bibfnamefont {E.-M.}\ \bibnamefont {Laine}},
  \bibinfo {author} {\bibfnamefont {J.}~\bibnamefont {Piilo}}, \ and\ \bibinfo
  {author} {\bibfnamefont {B.}~\bibnamefont {Vacchini}},\ }\href {\doibase
  10.1103/RevModPhys.88.021002} {\bibfield  {journal} {\bibinfo  {journal}
  {Rev. Mod. Phys.}\ }\textbf {\bibinfo {volume} {88}},\ \bibinfo {pages}
  {021002} (\bibinfo {year} {2016})}\BibitemShut {NoStop}%
\bibitem [{\citenamefont {Gorini}\ \emph {et~al.}(1976)\citenamefont {Gorini},
  \citenamefont {Kossakowski},\ and\ \citenamefont {Sudarshan}}]{Gorini1976a}%
  \BibitemOpen
  \bibfield  {author} {\bibinfo {author} {\bibfnamefont {V.}~\bibnamefont
  {Gorini}}, \bibinfo {author} {\bibfnamefont {A.}~\bibnamefont {Kossakowski}},
  \ and\ \bibinfo {author} {\bibfnamefont {E.~C.~G.}\ \bibnamefont
  {Sudarshan}},\ }\href@noop {} {\bibfield  {journal} {\bibinfo  {journal} {J.
  Math. Phys.}\ }\textbf {\bibinfo {volume} {17}},\ \bibinfo {pages} {821}
  (\bibinfo {year} {1976})}\BibitemShut {NoStop}%
\bibitem [{\citenamefont {Lindblad}(1976)}]{Lindblad1976a}%
  \BibitemOpen
  \bibfield  {author} {\bibinfo {author} {\bibfnamefont {G.}~\bibnamefont
  {Lindblad}},\ }\href@noop {} {\bibfield  {journal} {\bibinfo  {journal}
  {Comm. Math. Phys.}\ }\textbf {\bibinfo {volume} {48}},\ \bibinfo {pages}
  {119} (\bibinfo {year} {1976})}\BibitemShut {NoStop}%
\bibitem [{\citenamefont {Breuer}\ \emph {et~al.}(2009)\citenamefont {Breuer},
  \citenamefont {Laine},\ and\ \citenamefont {Piilo}}]{Breuer2009b}%
  \BibitemOpen
  \bibfield  {author} {\bibinfo {author} {\bibfnamefont {H.-P.}\ \bibnamefont
  {Breuer}}, \bibinfo {author} {\bibfnamefont {E.-M.}\ \bibnamefont {Laine}}, \
  and\ \bibinfo {author} {\bibfnamefont {J.}~\bibnamefont {Piilo}},\
  }\href@noop {} {\bibfield  {journal} {\bibinfo  {journal} {Phys. Rev. Lett.}\
  }\textbf {\bibinfo {volume} {103}},\ \bibinfo {pages} {210401} (\bibinfo
  {year} {2009})}\BibitemShut {NoStop}%
\bibitem [{\citenamefont {Rivas}\ \emph {et~al.}(2010)\citenamefont {Rivas},
  \citenamefont {Huelga},\ and\ \citenamefont {Plenio}}]{Rivas2010a}%
  \BibitemOpen
  \bibfield  {author} {\bibinfo {author} {\bibfnamefont {A.}~\bibnamefont
  {Rivas}}, \bibinfo {author} {\bibfnamefont {S.~F.}\ \bibnamefont {Huelga}}, \
  and\ \bibinfo {author} {\bibfnamefont {M.~B.}\ \bibnamefont {Plenio}},\
  }\href {http://journals.aps.org/prl/abstract/10.1103/PhysRevLett.105.050403}
  {\bibfield  {journal} {\bibinfo  {journal} {Phys. Rev. Lett.}\ }\textbf
  {\bibinfo {volume} {105}},\ \bibinfo {pages} {050403} (\bibinfo {year}
  {2010})}\BibitemShut {NoStop}%
\bibitem [{\citenamefont {Chruscinski}\ \emph {et~al.}(2011)\citenamefont
  {Chruscinski}, \citenamefont {Kossakowski},\ and\ \citenamefont
  {Rivas}}]{Chruscinski2011a}%
  \BibitemOpen
  \bibfield  {author} {\bibinfo {author} {\bibfnamefont {D.}~\bibnamefont
  {Chruscinski}}, \bibinfo {author} {\bibfnamefont {A.}~\bibnamefont
  {Kossakowski}}, \ and\ \bibinfo {author} {\bibfnamefont {A.}~\bibnamefont
  {Rivas}},\ }\href {\doibase 10.1103/PhysRevA.83.052128} {\bibfield  {journal}
  {\bibinfo  {journal} {Phys. Rev. A}\ }\textbf {\bibinfo {volume} {83}},\
  \bibinfo {pages} {052128} (\bibinfo {year} {2011})}\BibitemShut {NoStop}%
\bibitem [{\citenamefont {Wi\ss{}mann}\ \emph {et~al.}(2015)\citenamefont
  {Wi\ss{}mann}, \citenamefont {Breuer},\ and\ \citenamefont
  {Vacchini}}]{Wissmann2015a}%
  \BibitemOpen
  \bibfield  {author} {\bibinfo {author} {\bibfnamefont {S.}~\bibnamefont
  {Wi\ss{}mann}}, \bibinfo {author} {\bibfnamefont {H.-P.}\ \bibnamefont
  {Breuer}}, \ and\ \bibinfo {author} {\bibfnamefont {B.}~\bibnamefont
  {Vacchini}},\ }\href {\doibase 10.1103/PhysRevA.92.042108} {\bibfield
  {journal} {\bibinfo  {journal} {Phys. Rev. A}\ }\textbf {\bibinfo {volume}
  {92}},\ \bibinfo {pages} {042108} (\bibinfo {year} {2015})}\BibitemShut
  {NoStop}%
\bibitem [{\citenamefont {Amato}\ \emph {et~al.}(2018)\citenamefont {Amato},
  \citenamefont {Breuer},\ and\ \citenamefont {Vacchini}}]{Amato2018a}%
  \BibitemOpen
  \bibfield  {author} {\bibinfo {author} {\bibfnamefont {G.}~\bibnamefont
  {Amato}}, \bibinfo {author} {\bibfnamefont {H.-P.}\ \bibnamefont {Breuer}}, \
  and\ \bibinfo {author} {\bibfnamefont {B.}~\bibnamefont {Vacchini}},\ }\href
  {\doibase 10.1103/PhysRevA.98.012120} {\bibfield  {journal} {\bibinfo
  {journal} {Phys. Rev. A}\ }\textbf {\bibinfo {volume} {98}},\ \bibinfo
  {pages} {012120} (\bibinfo {year} {2018})}\BibitemShut {NoStop}%
\bibitem [{\citenamefont {Gardiner}\ and\ \citenamefont
  {Zoller}(2000)}]{Gardiner2000a}%
  \BibitemOpen
  \bibfield  {author} {\bibinfo {author} {\bibfnamefont {C.~W.}\ \bibnamefont
  {Gardiner}}\ and\ \bibinfo {author} {\bibfnamefont {P.}~\bibnamefont
  {Zoller}},\ }\href@noop {} {\emph {\bibinfo {title} {Quantum Noise}}}\
  (\bibinfo  {publisher} {Springer},\ \bibinfo {address} {New York},\ \bibinfo
  {year} {2000})\BibitemShut {NoStop}%
\bibitem [{\citenamefont {Khalfin}(1958)}]{Khalfin1958a}%
  \BibitemOpen
  \bibfield  {author} {\bibinfo {author} {\bibfnamefont {L.~A.}\ \bibnamefont
  {Khalfin}},\ }\href@noop {} {\bibfield  {journal} {\bibinfo  {journal}
  {JETP}\ }\textbf {\bibinfo {volume} {33}},\ \bibinfo {pages} {1371} (\bibinfo
  {year} {1958})}\BibitemShut {NoStop}%
\bibitem [{\citenamefont {Alicki}(2002)}]{Alicki2002b}%
  \BibitemOpen
  \bibfield  {author} {\bibinfo {author} {\bibfnamefont {R.}~\bibnamefont
  {Alicki}},\ }in\ \href@noop {} {\emph {\bibinfo {booktitle} {Dynamical
  semigroups: Dissipation, chaos, quanta}}},\ \bibinfo {series} {Lecture Notes
  in Physics}, Vol.\ \bibinfo {volume} {597},\ \bibinfo {editor} {edited by\
  \bibinfo {editor} {\bibfnamefont {P.}~\bibnamefont {Garbaczewski}}\ and\
  \bibinfo {editor} {\bibfnamefont {R.}~\bibnamefont {Olkiewicz}}}\ (\bibinfo
  {publisher} {Springer-Verlag},\ \bibinfo {address} {Berlin},\ \bibinfo {year}
  {2002})\ pp.\ \bibinfo {pages} {239--264}\BibitemShut {NoStop}%
\bibitem [{\citenamefont {Davies}(1974)}]{Davies1974a}%
  \BibitemOpen
  \bibfield  {author} {\bibinfo {author} {\bibfnamefont {E.~B.}\ \bibnamefont
  {Davies}},\ }\href {http://projecteuclid.org/euclid.cmp/1103860160}
  {\bibfield  {journal} {\bibinfo  {journal} {Comm. Math. Phys.}\ }\textbf
  {\bibinfo {volume} {39}},\ \bibinfo {pages} {91} (\bibinfo {year}
  {1974})}\BibitemShut {NoStop}%
\bibitem [{\citenamefont {Gorini}\ and\ \citenamefont
  {Kossakowski}(1976)}]{Gorini1976b}%
  \BibitemOpen
  \bibfield  {author} {\bibinfo {author} {\bibfnamefont {V.}~\bibnamefont
  {Gorini}}\ and\ \bibinfo {author} {\bibfnamefont {A.}~\bibnamefont
  {Kossakowski}},\ }\href {\doibase 10.1063/1.523057} {\bibfield  {journal}
  {\bibinfo  {journal} {J. Math. Phys.}\ }\textbf {\bibinfo {volume} {17}},\
  \bibinfo {pages} {1298} (\bibinfo {year} {1976})}\BibitemShut {NoStop}%
\bibitem [{\citenamefont {D{\"u}mcke}(1985)}]{Dumcke1985a}%
  \BibitemOpen
  \bibfield  {author} {\bibinfo {author} {\bibfnamefont {R.}~\bibnamefont
  {D{\"u}mcke}},\ }\href@noop {} {\bibfield  {journal} {\bibinfo  {journal}
  {Commun. Math. Phys.}\ }\textbf {\bibinfo {volume} {97}},\ \bibinfo {pages}
  {331} (\bibinfo {year} {1985})}\BibitemShut {NoStop}%
\bibitem [{\citenamefont {Hudson}\ and\ \citenamefont
  {Parthasarathy}(1984)}]{Hudson1984a}%
  \BibitemOpen
  \bibfield  {author} {\bibinfo {author} {\bibfnamefont {R.~L.}\ \bibnamefont
  {Hudson}}\ and\ \bibinfo {author} {\bibfnamefont {K.~R.}\ \bibnamefont
  {Parthasarathy}},\ }\href
  {https://projecteuclid.org:443/euclid.cmp/1103941122} {\bibfield  {journal}
  {\bibinfo  {journal} {Comm. Math. Phys.}\ }\textbf {\bibinfo {volume} {93}},\
  \bibinfo {pages} {301} (\bibinfo {year} {1984})}\BibitemShut {NoStop}%
\bibitem [{\citenamefont {Barchielli}(2006)}]{Barchielli2006a}%
  \BibitemOpen
  \bibfield  {author} {\bibinfo {author} {\bibfnamefont {A.}~\bibnamefont
  {Barchielli}},\ }in\ \href@noop {} {\emph {\bibinfo {booktitle} {Open quantum
  systems. III}}},\ \bibinfo {series} {Lecture Notes in Math.}, Vol.\ \bibinfo
  {volume} {1882}\ (\bibinfo  {publisher} {Springer},\ \bibinfo {address}
  {Berlin},\ \bibinfo {year} {2006})\ pp.\ \bibinfo {pages}
  {207--292}\BibitemShut {NoStop}%
\bibitem [{\citenamefont {Barchielli}\ and\ \citenamefont
  {Gregoratti}(2013)}]{Barchielli2013a}%
  \BibitemOpen
  \bibfield  {author} {\bibinfo {author} {\bibfnamefont {A.}~\bibnamefont
  {Barchielli}}\ and\ \bibinfo {author} {\bibfnamefont {M.}~\bibnamefont
  {Gregoratti}},\ }\href@noop {} {\bibfield  {journal} {\bibinfo  {journal}
  {Quantum Meas. Quantum Metrol.}\ }\textbf {\bibinfo {volume} {1}},\ \bibinfo
  {pages} {34 } (\bibinfo {year} {2013})}\BibitemShut {NoStop}%
\bibitem [{\citenamefont {Barchielli}\ and\ \citenamefont
  {Vacchini}(2015)}]{Barchielli2015a}%
  \BibitemOpen
  \bibfield  {author} {\bibinfo {author} {\bibfnamefont {A.}~\bibnamefont
  {Barchielli}}\ and\ \bibinfo {author} {\bibfnamefont {B.}~\bibnamefont
  {Vacchini}},\ }\href@noop {} {\bibfield  {journal} {\bibinfo  {journal} {New
  J.~Phys.}\ }\textbf {\bibinfo {volume} {17}},\ \bibinfo {pages} {083004}
  (\bibinfo {year} {2015})}\BibitemShut {NoStop}%
\bibitem [{\citenamefont {Ziman}\ \emph {et~al.}(2005)\citenamefont {Ziman},
  \citenamefont {{\v S}telmachovi{\v c}},\ and\ \citenamefont {Bu{\v
  z}ek}}]{Ziman2005a}%
  \BibitemOpen
  \bibfield  {author} {\bibinfo {author} {\bibfnamefont {M.}~\bibnamefont
  {Ziman}}, \bibinfo {author} {\bibfnamefont {P.}~\bibnamefont {{\v
  S}telmachovi{\v c}}}, \ and\ \bibinfo {author} {\bibfnamefont
  {V.}~\bibnamefont {Bu{\v z}ek}},\ }\href {\doibase 10.1007/s11080-005-0488-0}
  {\bibfield  {journal} {\bibinfo  {journal} {Open Syst. Inf. Dyn.}\ }\textbf
  {\bibinfo {volume} {12}},\ \bibinfo {pages} {81} (\bibinfo {year}
  {2005})}\BibitemShut {NoStop}%
\bibitem [{\citenamefont {Ciccarello}(2017)}]{Ciccarello2017a}%
  \BibitemOpen
  \bibfield  {author} {\bibinfo {author} {\bibfnamefont {F.}~\bibnamefont
  {Ciccarello}},\ }\href {http://stacks.iop.org/1402-4896/2013/i=T153/a=014010}
  {\bibfield  {journal} {\bibinfo  {journal} {Quantum Meas. Quantum Metrol.}\
  }\textbf {\bibinfo {volume} {4}},\ \bibinfo {pages} {53} (\bibinfo {year}
  {2017})}\BibitemShut {NoStop}%
\end{thebibliography}

%

\end{document}